\documentclass{aa}
\usepackage{graphicx}
\usepackage[varg]{txfonts}
\usepackage{natbib}
\bibpunct{(}{)}{; }{a}{}{,} % to follow the A&A style
\usepackage{multirow}
\usepackage{booktabs}
\begin{document}

\title{A statistical method for the identification of stars enriched in neutron-capture elements from medium-resolution spectra}

\author{G.\ Nav\'o
  \inst{1,2}
  \and
  J.~L.\ Tous
  \inst{1,2}
  \and
  J.~M.\ Solanes
  \inst{1,2}}
\institute{Departament de F\'{\i}sica Qu\`antica i Astrof\'{\i}sica, Universitat de
  Barcelona. C.\ Mart\'{\i} i Franqu\`es, 1; E--08028~Barcelona, Spain
  \and
  Institut de Ci\`encies del Cosmos (ICCUB), Universitat de
  Barcelona. C.\ Mart\'{\i} i Franqu\`es, 1; E--08028~Barcelona, Spain
  }
\offprints{G.\ Nav\'o, \email{gerrinp@gmail.com}}

% These dates will be filled out by the publisher
%\date{Submitted to A\&A. Revised version no.\ 3. Please, do not circulate.} 
% slugcomment
\date{Accepted by \aap.}
\graphicspath{{figs/}}

\abstract{We present an automated statistical method that uses medium-resolution spectroscopic observations of a set of stars to select those that show evidence of possessing significant amounts of neutron-capture elements. Our tool was tested against a sample of $\sim 70,000$ F- and G-type stars distributed among $215$ plates from the Galactic Understanding and Exploration (SEGUE) survey, including $13$ that were directed at stellar Galaxy clusters. Focusing on five spectral lines of europium in the visible window, our procedure ranked the stars by their likelihood of having enhanced content of this atomic species and identifies the objects that exhibit signs of being rich in neutron-capture elements as those scoring in the upper $2.5\%$. We find that several of the cluster plates contain relatively large numbers of stars with significant absorption around at least three of the five selected lines. The most prominent is the globular cluster M3, where we measured a fraction of stars that are potentially rich in heavy nuclides, representing at least $15\%$.}


\titlerunning{Identifying stars enriched in n-capture elements}
\authorrunning{G.\ Nav\'o et al.}

\maketitle

%\tableofcontents

\section{Introduction}

Metals in stars today are a snapshot of the metals in the interstellar medium (ISM) at the time and place where stars were born. While element synthesis is reasonably well understood from the primordial lightest elements up to the iron peak via fusion reactions that take place in the stellar interiors, explaining the origin of elements that are heavier than iron remains one of the major challenges in modern astrophysics. In order to accomplish this, it is essential to gather detailed and accurate information on the stellar abundances of the elements in question.

Since the binding energy per nucleon only increases until $^{62}$Ni (though it is generally believed that $^{56}$Fe is more common than nickel isotopes), heavier elements present in ancient halo stars, the ISM, dust grains, meteorites, and on Earth cannot be produced in a fusion process and must form by reactions of neutron (n) capture usually followed by a $\beta^{-}$ decay (most commonly) or $\beta^{+}$ decay, which have the generic form:
\begin{equation}
    \label{neutron capture}
    ^A_ZX + n\ \rightarrow\  ^{A+1}_ZX + \gamma\,,
\end{equation}
\begin{equation}
    \label{beta decay -}
   ^{A+1}_ZX \rightarrow\ ^{A+1}_{Z+1}X + e^{-} + \bar{\nu}_{e}\qquad\ \ \ \ \ \ \mbox{if}\ \ \beta^{-} \,,
\end{equation}
or
\begin{equation}
    \label{beta decay +}
    ^{A+1}_ZX \rightarrow\ ^{A+1}_{Z-1}X + e^{+} + \nu_{e}\qquad\ \ \ \ \ \ \mbox{if}\ \ \beta^{+} \,, \tag{\theequation${}^\prime$}
\end{equation}
where $A$ and $Z$ are the mass number and the atomic number of the nuclide, respectively.

The two predominant n-capture mechanisms are the s-process and the r-process, which are from slow and rapid neutron capture, respectively \citep{burbi,meyer,sneden03}. In s-process reactions, the characteristic time of $\beta$-decay, $\tau_{\beta}$ (Eqs.\ [\ref{beta decay -}] and [\ref{beta decay +}]), is short compared to the neutron capture time (Eq.\ [\ref{neutron capture}]), $\tau_{n}$, that is $\tau_{\beta}<<\tau_{n}$, creating elements close to the floor of the valley of stability, which is formed by long-lived heavy nuclides \citep{busso99,sneden08,heil,kappe}. Instead, in r-process reactions the time related to $\beta$-decay is large compared to the neutron capture time, $\tau_{\beta}>>\tau_{n}$, allowing the capture of several neutrons before the nuclei have time to undergo radioactive decay \citep{rene,thiele11,arno}. As a result, the r-process typically synthesizes the heaviest isotopes of every heavy element. 

\begin{figure*}[t]
\includegraphics{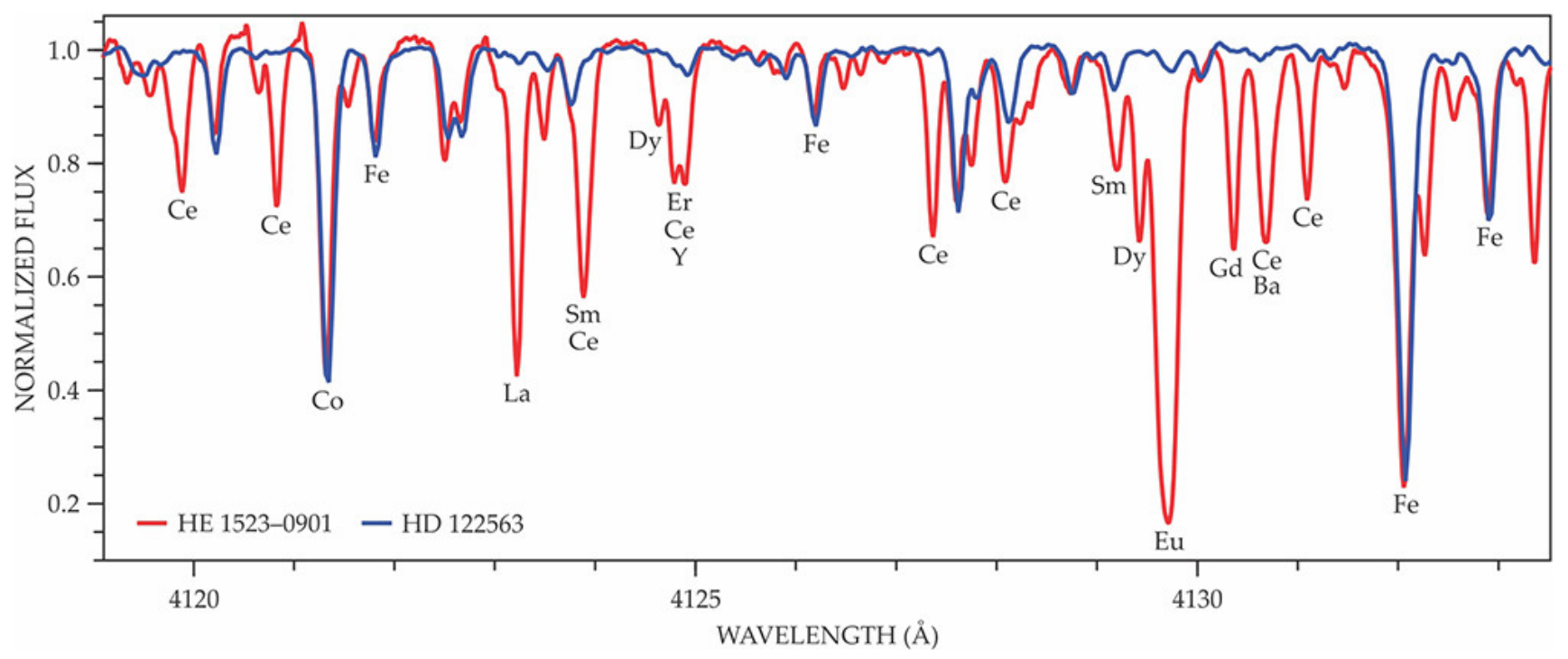}
\caption{High-resolution spectra of two stars: HD 122563 (blue) and HE 1523-0901 (red). While the former star has low abundances of r-process elements, the latter is the richest object in these elements ever found in the Galaxy's halo. It is important to note that the spectrum of HE 1523-0901 shows a large number of strong absorption lines from lanthanides such as lanthanum (La), cerium (Ce), samarium (Sm) and europium (Eu), which produces the most intense absorption line at $4129.70$\thinspace\AA. \citetext{Reproduced from \citealt{frebel}, with the permission of the American Institute of Physics.}
}
\label{fig:wide}
\end{figure*}

While $\tau_\beta$ only depends on the nuclear species, $\tau_{n}$ hinges on the neutron flux involved. The s-process takes place within stars, particularly during He-burning. Therefore, it can occur in the He-shell flashes of asymptotic giant branch (AGB) stars or in the He-burning cores of massive stars ($M>12\ \mbox{M}_\odot$) where new heavy elements are synthesized through the capture of free neutrons by preexisting heavy nuclei. This mostly arises in iron isotopes that are left, for instance, by a previous stellar generation \citep{lugaro,karak,battis,gorie,limon}. Thus, the observation of an overabundance of s-process elements in low-mass halo stars that cannot produce them is nowadays explained by the transfer from a more massive companion, currently a white dwarf, that has gone through some of these He-burning phases \citetext{see e.g., the review by \citealt{sneden08} and references therein}. Due to the relatively low neutron densities involved in this process \citetext{$\sim 10^8\,\mbox{cm}^{-3}$; \citealt{busso99}}, typically several decades can pass between consecutive neutron captures. 

On the other hand, although the astrophysical sources of the r-process are still not definitively identified, it is well known that this is a primary process that only occurs in environments with extremely large neutron densities \citetext{$\sim 10^{20}$--$10^{28}\,\mbox{cm}^{-3}$ depending on the source consulted, e.g., \citealt{kratz,frebel,licca}}, meaning that it cannot occur in stellar interiors. The huge densities involved also imply, contrary to the s-process, that the r-process can itself create free nucleons as well as the heavy seeds (iron-peak nuclei) that are required to build heavier elements.

Traditionally, the most probable site for the r-process has been assumed to be the large regions swept by the ejecta (i.e.,\ neutrino winds) of core-collapse supernovae \citetext{SNII; \citealt{burbi,woosley1,cowan,thiele18}}. However, theoretical models that rely on this mechanism have difficulties in explaining the observed, very low abundance of r-process nuclides in the interstellar medium, as well as the required high density of free neutrons in the ejected material. 

Another possible, presumably less uniformly distributed, source of neutron-rich matter are binary neutron star (NS-NS) mergers \citep{freiburghaus,komiya16,kasen,naiman}. The feasibility of this scenario was recently confirmed in 2017 when the gravitational waves coming from the LIGO/Virgo Event GW170817 were used to identify the location of a kilonova\footnote{A rapidly fading optical-infrared transient powered by the radioactive decay of n-rich species synthesized in NS-NS mergers.} whose rapidly expanding ejecta revealed spectral features consistent with the presence of high-opacity lanthanides \citep{pian,arcavi,smartt}. Assuming the solar r-process abundance pattern for the ejecta, current Galactic chemical evolution models suggest that this NS-NS merger event may have generated between 1--5 Earth masses of europium (Eu) and 3--13 Earth masses of gold \citep{argast,cote}. This is a much higher yield of heavy materials than in typical SNII, which compensates for their alleged greater rarity. More importantly, the range of the cosmic NS-NS merger rate estimated by LIGO/Virgo, although still poorly constrained, is consistent with the range required by these models to explain the nucleosynthesis of heavy elements in the Milky Way. 

Very recently, collapsars -- the supernova-triggering collapse of rapidly spinning, massive stars into black holes -- have also been suggested to explain the presence of heavy elements in the stars that formed early in the history of the universe. This rare type of supernovae can occur shortly after the first stars begin to form and they are expected to be even more prolific producers of r-process elements than NS-NS mergers \citep{siegel}. All of these findings are shifting the focus from SNII to both binary NS mergers and collapsars as the primary r-process sites in the universe.

The aim of this work is to provide a tool to locate regions of our Galaxy that show signs of being rich in n-capture elements, such as those arising from the r-process mentioned above. For this purpose, we use the complete spectral sample from the Sloan Extension for Galactic Understanding and Exploration (SEGUE) survey collected under the program names segue2 and segcluster during the first and second extensions of the Sloan Digital Sky Survey that was made available for its eighth Data Release \citetext{SDSS-III DR8; \citealt{segue,sdss3}}. The segue2 program plates contain spectra for approximately $119,000$ stars, and the program focuses on the in situ stellar halo of the Galaxy at distances between $10-60$ kpc. Among the different target types included in this survey, there are blue horizontal branch stars, F and G main-sequence stars, and K- and M-giants. On the other hand, the segcluster program consists of 13 plates specifically targeted on known globular or open clusters. We aim to search for the presence of neutral or singly ionized Eu in the spectra of these stars since this element is one of the best indicators of enrichment by rapid n-capture reactions \citep{argast}. In particular, Eu is an element synthesized practically on its totality through the r-process \citetext{$97\%$ at solar metallicity according to \citealt{burris}} that has several strong absorption lines in the part of the electromagnetic spectrum falling within the wavelength coverage of SEGUE. Fig.~\ref{fig:wide} illustrates the major differences that can result in a small range of visible wavelengths between an n-capture-rich star and one that is not when they are observed with a very high-resolution spectrometer. 

This article is organized as follows. In Section \ref{Sec:methodology} we describe our data sample and the analysis performed. In Section \ref{Sec:Results} we briefly discuss the results obtained. Finally, in Section \ref{Sec:Conclusions} we present the main conclusions of this work.

\section{Methodology}
\label{Sec:methodology}

\subsection{The data}

The spectra range of SEGUE is from $385$ nm to $920$ nm, with an average spectral resolution of $R\equiv \lambda /\Delta \lambda\sim 2,000$ (4 km s$^{-1}$).\ Although it does not allow one to resolve the usually very thin lines corresponding to atomic electronic transitions, it is, however, good enough to make the application of the statistical procedure that we describe in this article feasible. The SEGUE sample consists primarily of two surveys, SEGUE-1 and SEGUE-2, which in turn encompass several programs that vary in their observational focus. Each program consists of a number of different plates or tiles, each covering an area of 7 ${\rm deg}^{2}$ and containing $640$ fibers. The segue2 program mainly focuses on the distant northern Galactic halo region. It consists of $202$ tiles that encompass a total sky area of $1,317\thinspace{\rm deg}^{2}$ and spectroscopically observe a total of $118,151$ unique stars up to a magnitude of $g=19$. The segue2 "blind" plates contain stars that are likely heterogeneous in terms of their chemical abundances because of the mixing that results from the high proper motions, which are characteristic of this structural component. In contrast, the segcluster program focuses on regions dominated by open and globular clusters, that is, on collections of stars that are expected to be more chemically homogeneous because they contain numerous objects that have been gravitationally bound since the day they were born. 

In this work, we analyze the subset of high-quality, high signal-to-noise ratio (S/N) spectra (see next section) of $\sim 70,000$ stars of F and G spectral types. In fact, the adopted dataset is heavily biased against G-type stars. Therefore, we were forced to limit the application of our statistical procedure to the warmest objects (subclasses G0--G2) of G-type stars. This constraint was far from an inconvenience and was actually beneficial for this analysis since we excluded the coolest G-type objects, which are also the most likely to be affected by absorption lines from diatomic molecules (CH, CN) in the spectral region studied. Depending on the intensity of these lines, they could potentially lead to false detections, something that might happen if our method is applied to the colder K- and M-type stars. 

\subsection{Spectral analysis}
\label{SSec:spectral_analysis}

Our methodology essentially consists of first averaging the spectra for all of the stars with the same spectral type and then comparing each individual spectrum with the inferred mean to identify the most discordant ones (i.e., the most flux deficient) in several narrow spectral bands that encompass certain strong absorption lines of n-capture elements. The averaging process is based on the algorithm described in detail by \citeauthor{Mas} \citetext{\citeyear{Mas}; see references therein}, which we modified in order to adapt it to our needs. We began by shifting each spectrum to the laboratory rest-frame. Because the binning in wavelength has a constant logarithmic dispersion, the correction is given by $\log \lambda_{\mathrm r} = \log \lambda_{\mathrm o} - \log (1 + z)$, where $\lambda_{\mathrm r}$ is the rest-frame wavelength and $\lambda_{\mathrm o}$ is the observed one. After this correction, we rebinned the flux and its error by interpolating into pixels with a constant logarithmic spacing of $0.0001$, so the resolution of the original spectra was preserved along the full visible range\footnote{\url{http://www.sdss.org/dr12/spectro/spectro_basics/}}. Next, we removed the pixels whose errors are set to infinity from the spectra as well as those that have the mask bit {\texttt{BRIGHTSKY}} activated\footnote{See SDSS bitmasks at  \url{http://www.sdss.org/dr12/algorithms/bitmasks/\#SPPIXMASK}.}. Any spectrum containing more than $50\%$ of problematic pixels was  discarded from the sample. Furthermore, we also discarded from the analysis those spectra in which the most immediate neighborhood of the lines selected to probe the presence of n-rich species (see below) was not free of anomalies. This neighborhood is formed by the twenty pixels located to the left and right of the central pixel of these lines (i.e., the interval $\lambda_{i-20} \leq \lambda_i \leq \lambda_{i + 20} $, with $\lambda_i$ the pixel containing the wavelength of any line of interest). 

Then, we used the tool iSpec (\citealt{iSpec1, iSpec2}) for each spectrum, $j$, that fit the criteria just mentioned to determine both its S/N, $s_j$, and normalized flux, $f_j$ (for the continuum normalization we used a spline of second degree every $5$ nm that ignores any region with strong lines). Finally, the normalized spectra having an S/N$\;\geqslant 10$ were stacked together as the weighted average
\begin{equation}
    \bar{f_{i}} = \dfrac{\sum_{j} s_j\cdot f_{ij}}{\sum_{j} s_j}\;,
    \label{eq:mean_flux}
\end{equation}
where $\bar{f_{i}}$ is the mean normalized flux at pixel $i$. The error assigned to the mean fluxes is simply the sample standard deviation of the composite spectrum at each pixel:
\begin{equation}
\sigma_{\bar{f},i} = \sqrt{\dfrac{\sum_j \left(f_{ij} - \bar{f_i}\right)^2\cdot s_j}{\sum_j s_j}}\;.
\label{eq:mean_flux_err}
\end{equation}

To minimize the chances of making erroneous identifications, we chose a total of five lines of two different species of the r-process element Eu that are among the strongest lines of this element in the visible window. There are three intense lines of \ion{Eu}{I}, as well as two \ion{Eu}{II} resonance lines that are also among the most intense of this species, all of them located near the UV-Blue end of this window \citetext{see \citealt{nist} and references therein}. In order to detect the lines, we have taken into account that SEGUE's spectra have been obtained using a medium-resolution instrument with a spectral resolution of about two angstroms in the range of wavelengths of interest. This means that, in practice, the fluxes of the narrow, sub-angstrom-wide lines produced by heavy nuclides (see Fig.\ \ref{fig:wide}) end up distributed among several pixels around the peak\footnote{Pixel size at $ \sim 4000$\thinspace \AA\ is a bit less that one \AA.}.\ This makes them both wider and weaker and increases the likelihood that they would blend with adjacent lines. The absorption lines that we wish to detect are expected, therefore, to show up in the SDSS data as moderately deep depressions of the continuum that spread over a few angstroms. The different absorption lines adopted and the spectral ranges assigned for their detection are both listed in Table\ref{tab:species}.

\begin{figure}[!htb]
\centering
    \includegraphics[width=\columnwidth]{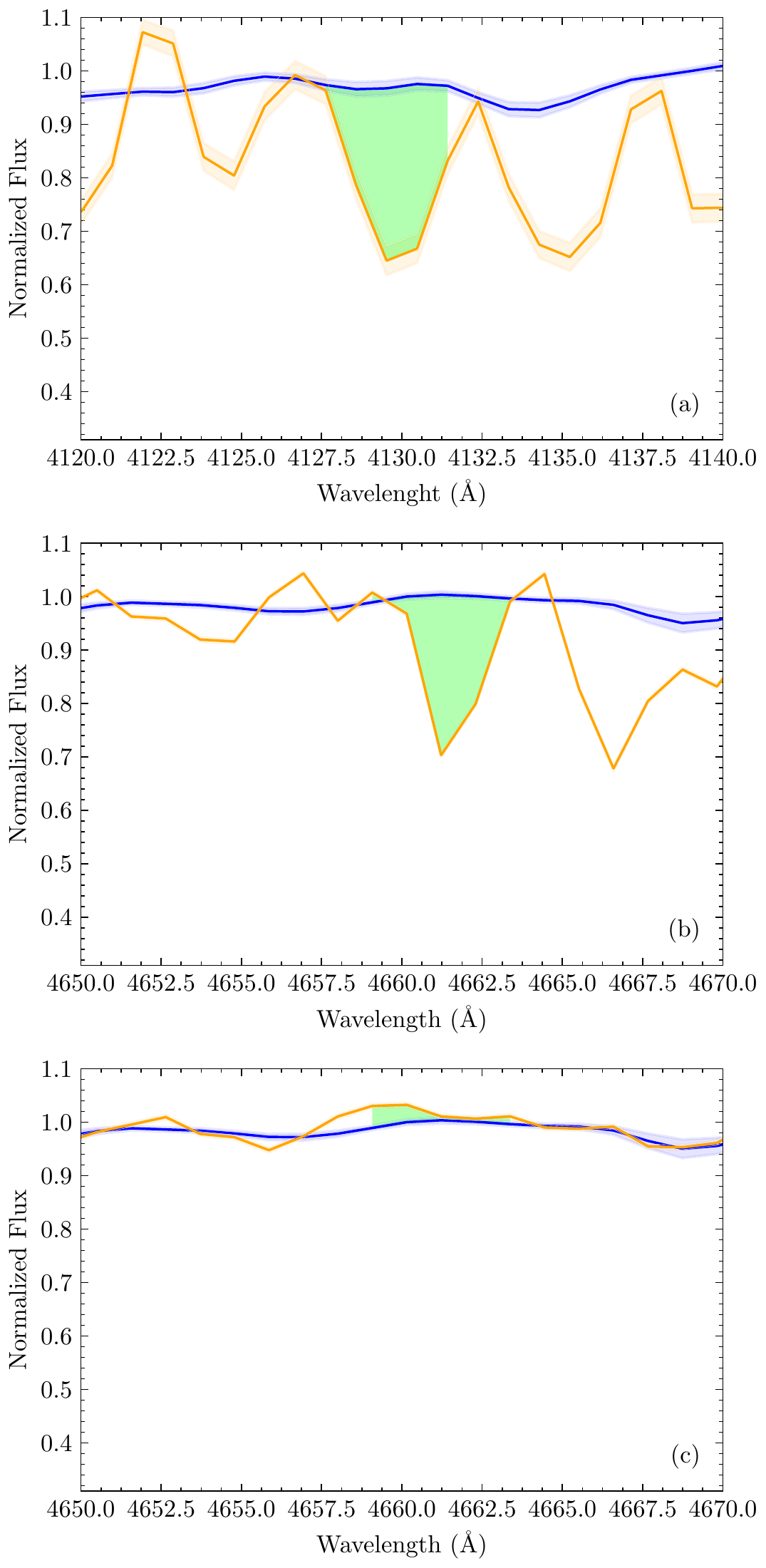}
%    \addlinespace[0.7cm]
   \caption{Comparison of certain observed lines in F5 stars (orange curves) with average flux inferred from whole spectral class (blue curves). The lighter bands surrounding the fluxes show their associated uncertainties, while the green colored areas identify the integrated flux differences that we attribute to lines from Eu. Panel (a) is for an F5 star ({\texttt{specobjID}} $= 3719984370237415424$) that shows a substantial flux deficit compared to the mean in the $4127$--$4131$ \AA\ wavelength range that encompasses an important line of \ion{Eu}{II} (see Table~\ref{tab:species}). Panel (b) depicts another F5 star ({\texttt{specobjID}} $= 2786716958031191040$) also showing a significant flux deficiency in the $4659$--$4663$ \AA\ range, which is the interval of wavelengths where one expects to find a strong \ion{Eu}{I} line. In both cases the stars have been identified as positive detections. In contrast, panel (c) shows a third F5 star ({\texttt{specobjID}} $= 3002850987342928896$) that does not show evidence for the existence of this latter line in absorption, which classifies it as a non-detection.
}
\label{fig:fig2}
\end{figure}

With all of this in mind, we devised a strategy to find candidate stars rich in n-capture elements, which consists in computing the differences between the fluxes of each individual stellar spectrum and the average spectrum of the corresponding spectral class calculated within four-\AA-wide spectral ranges around the adopted lines (see Table~\ref{tab:species}). The differences for each spectral range were then ranked in increasing order and those included in the upper $2.5$ percentile were considered to be indicators of the existence of an enhancement in the corresponding species. We also imposed that for a star to be considered a potential candidate rich in n-capture elements, it must have produced positive results in the overabundance of these elements in the majority fraction of the selected spectral regions, that is, in at least three of the five\footnote{This number and the minimum rank for detection were adopted interdependently to avoid selecting too many or too few candidate stars. We verified, however, that the results of the present analysis remain essentially unchanged as long as the values of these two parameters are kept within reasonable bounds.} Eu lines. The final step consisted of identifying the SEGUE plates that contain the highest number of stars with these characteristics. The idea behind this procedure is that we expect metal enrichment, especially when it is driven by rapid n-capture in spatially localized events, to manifest itself clearly in specific regions of the Galaxy, such as the star clusters targeted by the segcluster program where most objects share a common origin. Likewise, we hope that finding evidence of localized enrichment in heavy nuclides will be more difficult in the segue2 plates, since they include Galactic halo field stars with disparate formation histories. 

At this point, it also has to be borne in mind that some of the electronic transitions that we aim to detect produce absorption lines that are known to be ill-suited for equivalent-width analysis. This is because they can blend with nearby lines from lighter metal species and/or by molecular band regions that also lead to difficulties in placing the continuum \citep[e.g.,][]{koch,roeder,sique,battis}. Certainly, these factors can make the detection of such lines in individual stellar analyses very challenging (even from very high-resolution and S/N observations), which critically limits their usefulness as estimators of the abundance of n-capture elements. We want to emphasize, however, that the effects of crowding are taken into account in our methodology. This is so because it relies on the identification of the most statistically significant deviations in the fluxes from the spectral regions around lines of n-rich elements, which are self-consistently synthesized by averaging the spectra from thousands of similar Galactic halo stars that have been observed with the same instrument. This should also account for possible affectations related to hyperfine splitting due to the presence of different isotopes. In addition, the robustness of this procedure was boosted by introducing the requirement that, for a star to be considered an n-capture-rich candidate, it must present simultaneous indications of flux deficits in the mayority of the lines of interest. In any event, in an effort to be conservative, we subdivided our candidate stars into primary and secondary candidates, according to whether or not they show a significant flux deficit in the spectral ranges that are associated with at least one of the Eu absorption lines at 4129 \AA\  and 4594 \AA, which are two of the most frequently used indicators of the abundance of r-process elements in the visible spectral range because of their strength and lack of significant blending (see e.g., Fig.~\ref{fig:wide}).

\begin{table}[htb!]
\centering
\begin{tabular}{c c c}
\hline\hline
\addlinespace[0.05cm]
\textrm{Nuclide}&
\textrm{Spectral line (\AA)}&
\textrm{Spectral range (\AA)}\\
\hline
\addlinespace[0.05cm]
%\colrule
\multirow{3}{*}{\ion{Eu}{I}} & 4594 & 4592-4596\\
& 4627 & 4625-4629\\
& 4661 & 4659-4663\\ \hline
\addlinespace[0.05cm]
\multirow{2}{*}{\ion{Eu}{II}}  & 4129 & 4127-4131\\
& 4205 & 4203-4207\\ \hline
\addlinespace[0.1cm]
\end{tabular}
\caption{\label{tab:species}%
Strongest lines from Eu in optical window and assigned spectral ranges for detection (see text).}
\end{table} 

Figure \ref{fig:fig2}, provides a graphical example of the sort of spectral features with which our method operates. In the top and middle panels of the plot, we include two F5 stars whose stellar spectra show large deficits with respect to the average flux of this spectral class within the spectral ranges that delineate the \ion{Eu}{II} line at 4129 \AA\ and the \ion{Eu}{I} line at 4661 \AA, which are represented by solid green areas. In both cases the flux differences within the corresponding spectral ranges fall into the upper 2.5 percent and, therefore, are classified as potential detections of these species. This situation clearly contrasts with the bottom panel, where we show the spectrum of another F5 star that, instead of showing a significant decrease in the intensity within the spectral window associated with the \ion{Eu}{I} 4661 \AA\ line, presents a slight increase with respect to the mean, which classifies it as a non-detection. Next, we present and comment on the results obtained after applying our methodology to identify stars rich in n-capture elements to the collections of spectra gathered in the segue2 and segcluster samples.

\section{Results}
\label{Sec:Results}

The application of the statistical procedure described in the previous section led us to detect a total of 67 r-process-rich candidates in the different tiles that make up the segue2 and segcluster programs. Forty-three of these objects (39 primary and 4 secondary) are  distributed in the five plates that we include in Table~\ref{tab:results} where we list the plate ID (in Col.\ 2), the galactic coordinates of its center in degrees (Cols.\ 3 and 4), and the name of the targeted cluster (Col.\ 5) for those belonging to the segcluster program. These plates correspond to regions of the sky that include at least three positive detections, which is the minimum number required, in the present work, to consider that this part of the Galaxy would merit being explored in the search of n-capture enriched stars. As expected, all but one of these plates are devoted to star clusters. Besides, the only field halo (segue2) plate in this list contains the lowest total number of stars with an enhanced content in heavy elements in at least three lines, $N_{\geqslant 3}$ (Col.\ 6). Table~\ref{tab:results} also reports the number of stars per plate whose spectrum produces evidence that is favorable to the existence of significant amounts of elements arising from n-capture nucleosynthesis in exactly $l=3$, 4, and 5 of the narrow spectral ranges investigated and given by $N_3$, $N_4,$ and $N_5$ (Cols.\ 7--9)\footnote{$N_{\geqslant 3} = N_3 + N_4 + N_5$.}, respectively. The last two columns of Table~\ref{tab:results} provide information about the fraction of stars within each plate whose spectrum complies our quality requirements (see Sec.\ \ref{SSec:spectral_analysis}), which is defined by the ratio
\begin{equation}
    Q \equiv \frac{\text{\# high-quality spectra}}{\text{\# spectroscopic targets}}\;,
    \vspace{1 mm}
    \label{xi}
\end{equation}
and the frequency of candidate stars per plate
\begin{equation}
    \label{ri}
    F_{\mathrm r} \equiv \frac{\text{\# spectra rich in n-capture elements}}{\text{\# high-quality spectra}}\;.
    \vspace{1 mm}
\end{equation}

The comparison of the values reported for the last relation with its probability distribution function (PDF) inferred from the whole dataset allows one to observe their extreme nature: while the median of the PDF is $\bar{F}_{\mathrm r} = 0.0$ (and the mean is $\langle F_{\mathrm r}\rangle = 0.004$) all of the frequencies listed are well within its upper decile ($F_{\mathrm r} > 0.014$). Therefore, these figures can be considered as a piece of evidence that lends support to the claim that the selected plates are directed at locations in the Galaxy where there is an overabundance of stars rich in r-process elements. We note that four of these five plates sample areas of the sky containing collections of stars with a contemporary origin and that, therefore, are expected to have a similar chemical composition. On the other hand, the fact that the plates with the highest numbers of candidate stars (IDs 2475 and 2476) show $Q$ values very close to the average of the whole set, $\langle Q\rangle\simeq 0.2$, excludes the possibility that the number of these stars found in a plate depends on the fraction of spectra analyzed. It should be added that even if we only consider the candidate stars contained in these five plates, the probability of obtaining 43 positive signals from a sample of $70,000$ stars by chance is extremely low (in the order of $10^{-57}$, or $10^{-45}$ if we count the 39 primary detections)\footnote{These estimates are based on a binomial distribution with a probability $p$ of success (i.e., detecting three or more lines on a star by chance) equal to $1.6025\cdot 10^{-5}$, so the expected number of chance detections in a sample of $n\simeq 70,000$ stars would be $\sim 1\pm 1$.}. It is also important to note that eleven of the 43 candidate stars show indications of a heavy nuclide boost in four lines and two of them in all of the five lines (see Table~\ref{tab:results}). This reinforces the robustness of the detections. All of these very strong objective arguments confirm a posteriori the correct functioning of our method and, therefore, make us particularly confident in our identification, via SEGUE's spectra, of regions of our Galaxy that are potentially rich in n-capture-enhanced stars. The SDSS spectroscopic identifiers ({\texttt{specobjID}}), together with the spectral subclasses and metallicities \citetext{the latter are taken from \citealt{lee}} of all candidate stars enriched in Eu detected in the plates discussed above, can be found in Table\ \ref{tab:stars} of Appendix\ \ref{App:stars}.

\begin{table*}[!htb]
\centering
\begin{tabular}{l c r r r r r c c c c}
\hline\hline
\addlinespace[0.05cm]
\textrm{SEGUE program}&
\textrm{Plate ID}&
\multicolumn{1}{c}{$l\;(^{\circ}$)}&
\multicolumn{1}{c}{$b\;(^{\circ}$)}&
\textrm{Cluster ID}&
\textrm{$N_{\geqslant 3}$}&
\textrm{$N_3$}&
\textrm{$N_4$}&
\textrm{$N_5$}&
\textrm{$Q$}&
\textrm{$F_{\mathrm r}$}
\\
\hline
\addlinespace[0.05cm]
Segue2 & 3304 & 132.06 & 62.00 & --\ \ \ & 3 & 3 & 0 & 0 & 0.1 & 0.08\\ \hline
\addlinespace[0.05cm]
\multirow{4}{*}{Segcluster} & 2667 & 132.82 & 10.94 & M67 & 6 & 4 & 2 & 0 & 0.5 & 0.03\\
& 1961 & 54.00 & $-35.43$ & M2 & 6 & 4 & 1 & 1 & 0.3 & 0.03\\
& 2476 & 199.03 & 17.01 & NGC5053 & 10 & 8 & 2 & 0 & 0.3 & 0.07\\
& 2475 & 42.31 & 78.70 & M3 & 18 & 13 & 4 & 1 & 0.2 & 0.15\\
\hline
\addlinespace[0.1cm]
\end{tabular}
\caption{SEGUE plates with at least three candidate stars rich in n-capture elements.}
\label{tab:results}
\end{table*}

\section{Conclusions}
\label{Sec:Conclusions}

We have developed an automated statistical approach to identify stars containing enhanced contents of neutron-rich nuclei that use spectra whose resolution is not optimal to adequately detect from single observations the very thin absorption lines produced by this kind of elements. To illustrate how our methodology actually works, we focused on measuring the integrated flux within windows of a few angstrom wide around three lines of \ion{Eu}{I} and two of \ion{Eu}{II} that fall within the visible spectrum, subsequently identifying the strongest flux deficiencies found in these narrow spectral ranges with the presence of absorption features linked to these elements. For a star to be considered rich in heavy elements, we require it to test positive in at least three of the five lines investigated.

The performance of the procedure has been tested against a sample of around $70,000$ high-quality, medium-resolution optical spectra from F- and G-type stars observed within $215$ plates belonging to two programs of the SDSS/SEGUE survey. We find that there is only one plate among the 202 of the segue2 program that reaches the minimum required content of three candidate stars enriched in n-capture elements. In contrast, it is found that up to four of the 13 plates of the segcluster program contain between six and eighteen stars where n-rich nuclei are plentiful. These findings are consistent with the expectation that most targets included in the tiles of the latter program, which are specifically directed at both globular and open star clusters, should be stars created in a common formative environment; while the presumably disparate birth places of the objects observed in the segue2's tiles should contribute to blur any signature that might exist from past heavy-element nucleosynthesis events in the targeted regions of our Galaxy. For those readers interested in deciphering the site and origin of the n-capture processes, we provide a table listing the spectral SDSS identifiers and other basic information about the 43 candidate stars rich in r-process elements (distinguishing primary and secondary) that have been detected in these five plates. 

Highlights of the numerical results obtained include the ten or more objects that are potentially rich in heavy elements found in each one of the NGC5053 and M3 globular clusters, which are the two stellar systems observed in the segcluster program with the highest fractions of r-process enriched stars: $7\%$ and $15\%$, respectively. These results, however, exclusively report on the presence of stars with extreme photospheric abundances of heavy elements within their spectral class. Nevertheless, they do not indicate what could be the particular values of such abundances, neither in absolute terms nor normalized to the stars' metallicities, which are normally inferred from the Fe abundance. Since the latter, despite being subsolar in all cases, span quite a large range (from [Fe/H]$\;\simeq -0.45$ to $-2.5$; see Table~\ref{tab:stars}), it seems reasonable to expect that both the selected stars and their host systems can also end up showing substantial variations in the abundance ratios of heavy nuclides. Thus, on one end, only one of the six stars rich in r-process elements located in the region of the open star cluster M67 is truly metal-poor ([Fe/H]$\;\lesssim -1$). At the other end, there are the plates M2, NGC5053, and M3, which show fractions of objects with abundances of Fe below $10\%$ of 4/6, 8/10, and 9/10, respectively (for the last fraction we are considering only the stars with known metallic content), which would confirm that globular clusters are one of the best astrophysical sites to find stars with enhanced levels of n-capture elements. 

In summary, we devised a fast and reliable automated procedure that enables the detection of the fine absorption lines that are produced by heavy elements in stellar spectra by using data from large surveys whose spectral resolution is too coarse to approach this problem in a traditional way. Thus, our method can help, for instance, future high-resolution stellar spectral abundance surveys that search for clues as to the astrophysical origin of the r-process nuclides by facilitating the identification of the sites where it should be possible to find a good number of stars that exhibit enhancement in such elements. This would avoid the time-consuming searches through thousands of candidates that are currently being carried out to find these precious objects in the Galactic halo. Ultimately, this is about contributing to the collection of a more complete inventory of n-capture-enhanced stars that increases our understanding of the synthesis of heavy elements and that sheds more light on the formation history of galaxies from the chemical evolution of their stellar populations.

\section*{Acknowledgements}

The authors are grateful to an anonymous referee whose comments have prompted us to improve our methodology and its presentation, as well as better emphasize the reliability of the results. We also warmly thank Jaime Perea and Mar\'\i a Asunci\'on del Olmo for interesting discussions and useful advice about spectra normalization. J.L.T.\ and J.M.S.\ acknowledge financial support from the Spanish AEI and European FEDER funds through the research project AYA2016-76682-C3. Additional funding has been provided by the State Agency for Research of the Spanish MCIU through a 'Center of Excellence Mar\'\i a de Maeztu' award to the Institut de Ci\`encies del Cosmos of the University of Barcelona.

%\nocite{*}

\bibliography{36324_final}% Produces the bibliography via BibTeX.

\begin{thebibliography}{42}
\expandafter\ifx\csname natexlab\endcsname\relax\def\natexlab#1{#1}\fi

\bibitem[{{Arcavi} {et~al.}(2017){Arcavi}, {Hosseinzadeh}, {Howell}, {McCully},
  {Poznanski}, {Kasen}, {Barnes}, {Zaltzman}, {Vasylyev}, \& {Maoz}}]{arcavi}
{Arcavi}, I., {Hosseinzadeh}, G., {Howell}, D.~A., {et~al.} 2017, \nat, 551, 64

\bibitem[{{Argast} {et~al.}(2004){Argast}, {Samland}, {Thielemann}, \&
  {Qian}}]{argast}
{Argast}, D., {Samland}, M., {Thielemann}, F.-K., \& {Qian}, Y.-Z. 2004, \aap,
  416, 997

\bibitem[{{Arnould} {et~al.}(2007){Arnould}, {Goriely}, \& {Takahashi}}]{arno}
{Arnould}, M., {Goriely}, S., \& {Takahashi}, K. 2007, \physrep, 450, 97

\bibitem[{{Battistini} \& {Bensby}(2016)}]{battis}
{Battistini}, C. \& {Bensby}, T. 2016, \aap, 586, A49

\bibitem[{{Blanco-Cuaresma}(2019)}]{iSpec2}
{Blanco-Cuaresma}, S. 2019, \mnras, 486, 2075

\bibitem[{{Blanco-Cuaresma} {et~al.}(2014){Blanco-Cuaresma}, {Soubiran},
  {Heiter}, \& {Jofr{\'e}}}]{iSpec1}
{Blanco-Cuaresma}, S., {Soubiran}, C., {Heiter}, U., \& {Jofr{\'e}}, P. 2014,
  \aap, 569, A111

\bibitem[{Burbidge {et~al.}(1957)Burbidge, Burbidge, Fowler, \& Hoyle}]{burbi}
Burbidge, E.~M., Burbidge, G.~R., Fowler, W.~A., \& Hoyle, F. 1957, Rev. Mod.
  Phys., 29, 547

\bibitem[{{Burris} {et~al.}(2000){Burris}, {Pilachowski}, {Armand roff},
  {Sneden}, {Cowan}, \& {Roe}}]{burris}
{Burris}, D.~L., {Pilachowski}, C.~A., {Armand roff}, T.~E., {et~al.} 2000, The
  Astrophysical Journal, 544, 302

\bibitem[{{Busso} {et~al.}(1999){Busso}, {Gallino}, \& {Wasserburg}}]{busso99}
{Busso}, M., {Gallino}, R., \& {Wasserburg}, G.~J. 1999, \araa, 37, 239

\bibitem[{C{\^{o}}t{\'{e}} {et~al.}(2018)C{\^{o}}t{\'{e}}, Fryer, Belczynski,
  Korobkin, Chru{\'{s}}li{\'{n}}ska, Vassh, Mumpower, Lippuner, Sprouse,
  Surman, \& Wollaeger}]{cote}
C{\^{o}}t{\'{e}}, B., Fryer, C.~L., Belczynski, K., {et~al.} 2018, The
  Astrophysical Journal, 855, 99

\bibitem[{{Cowan} \& {Thielemann}(2004)}]{cowan}
{Cowan}, J.~J. \& {Thielemann}, F.-K. 2004, Physics Today, 57, 10.47

\bibitem[{{Eisenstein} {et~al.}(2011){Eisenstein}, {Weinberg}, {Agol},
  {Aihara}, {Allende Prieto}, {Anderson}, {Arns}, {Aubourg}, {Bailey}, \&
  {Balbinot}}]{sdss3}
{Eisenstein}, D.~J., {Weinberg}, D.~H., {Agol}, E., {et~al.} 2011, \aj, 142, 72

\bibitem[{{Frebel} \& {Beers}(2018)}]{frebel}
{Frebel}, A. \& {Beers}, T.~C. 2018, Phys. Today, 71, 30

\bibitem[{{Freiburghaus} {et~al.}(1999){Freiburghaus}, {Rosswog}, \&
  {Thielemann}}]{freiburghaus}
{Freiburghaus}, C., {Rosswog}, S., \& {Thielemann}, F.-K. 1999, \apjl, 525,
  L121

\bibitem[{{Goriely} \& {Siess}(2018)}]{gorie}
{Goriely}, S. \& {Siess}, L. 2018, \aap, 609, A29

\bibitem[{{Heil} {et~al.}(2009){Heil}, {Juseviciute}, {K{\"a}ppeler},
  {Gallino}, {Pignatari}, \& {Uberseder}}]{heil}
{Heil}, M., {Juseviciute}, A., {K{\"a}ppeler}, F., {et~al.} 2009, \pasa, 26,
  243

\bibitem[{{K{\"a}ppeler} {et~al.}(2011){K{\"a}ppeler}, {Gallino}, {Bisterzo},
  \& {Aoki}}]{kappe}
{K{\"a}ppeler}, F., {Gallino}, R., {Bisterzo}, S., \& {Aoki}, W. 2011, Reviews
  of Modern Physics, 83, 157

\bibitem[{{Karakas} {et~al.}(2012){Karakas}, {Garc{\'\i}a-Hern{\'a}ndez}, \&
  {Lugaro}}]{karak}
{Karakas}, A.~I., {Garc{\'\i}a-Hern{\'a}ndez}, D.~A., \& {Lugaro}, M. 2012,
  \apj, 751, 8

\bibitem[{{Kasen} {et~al.}(2017){Kasen}, {Metzger}, {Barnes}, {Quataert}, \&
  {Ramirez-Ruiz}}]{kasen}
{Kasen}, D., {Metzger}, B., {Barnes}, J., {Quataert}, E., \& {Ramirez-Ruiz}, E.
  2017, \nat, 551, 80

\bibitem[{{Koch} \& {Edvardsson}(2002)}]{koch}
{Koch}, A. \& {Edvardsson}, B. 2002, Astronomy and Astrophysics, 381, 500

\bibitem[{{Komiya} \& {Shigeyama}(2016)}]{komiya16}
{Komiya}, Y. \& {Shigeyama}, T. 2016, \apj, 830, 76

\bibitem[{{Kramida} {et~al.}(2018){Kramida}, {Ralchenko}, {Reader}, \& {NIST
  ASD Team}}]{nist}
{Kramida}, A., {Ralchenko}, Y., {Reader}, J., \& {NIST ASD Team}. 2018, NIST
  Atomic Spectra Database (version 5.6.1), \url{https://physics.nist.gov/asd}

\bibitem[{{Kratz} {et~al.}(2007){Kratz}, {Farouqi}, {Pfeiffer}, {Truran},
  {Sneden}, \& {Cowan}}]{kratz}
{Kratz}, K.-L., {Farouqi}, K., {Pfeiffer}, B., {et~al.} 2007, \apj, 662, 39

\bibitem[{{Lee} {et~al.}(2008){Lee}, {Beers}, {Sivarani}, {Allende Prieto},
  {Koesterke}, {Wilhelm}, {Re Fiorentin}, {Bailer-Jones}, {Norris}, {Rockosi},
  {Yanny}, {Newberg}, {Covey}, {Zhang}, \& {Luo}}]{lee}
{Lee}, Y.~S., {Beers}, T.~C., {Sivarani}, T., {et~al.} 2008, \aj, 136, 2022

\bibitem[{{Liccardo} {et~al.}(2018){Liccardo}, {Malheiro}, {Hussein},
  {Carlson}, \& {Frederico}}]{licca}
{Liccardo}, V., {Malheiro}, M., {Hussein}, M.~S., {Carlson}, B.~V., \&
  {Frederico}, T. 2018, European Physical Journal A, 54, 221

\bibitem[{{Limongi} \& {Chieffi}(2018)}]{limon}
{Limongi}, M. \& {Chieffi}, A. 2018, \apjs, 237, 13

\bibitem[{{Lugaro} {et~al.}(2003){Lugaro}, {Herwig}, {Lattanzio}, {Gallino}, \&
  {Straniero}}]{lugaro}
{Lugaro}, M., {Herwig}, F., {Lattanzio}, J.~C., {Gallino}, R., \& {Straniero},
  O. 2003, \apj, 586, 1305

\bibitem[{{Mas-Ribas} {et~al.}(2017){Mas-Ribas}, {Miralda-Escud{\'e}},
  {P{\'e}rez-R{\`a}fols}, {Arinyo-i-Prats}, {Noterdaeme}, {Petitjean},
  {Schneider}, {York}, \& {Ge}}]{Mas}
{Mas-Ribas}, L., {Miralda-Escud{\'e}}, J., {P{\'e}rez-R{\`a}fols}, I., {et~al.}
  2017, \apj, 846, 4

\bibitem[{{Meyer}(1994)}]{meyer}
{Meyer}, B.~S. 1994, \araa, 32, 153

\bibitem[{{Naiman} {et~al.}(2018){Naiman}, {Pillepich}, {Springel},
  {Ramirez-Ruiz}, {Torrey}, {Vogelsberger}, {Pakmor}, {Nelson}, {Marinacci},
  {Hernquist}, {Weinberger}, \& {Genel}}]{naiman}
{Naiman}, J.~P., {Pillepich}, A., {Springel}, V., {et~al.} 2018, Monthly
  Notices of the Royal Astronomical Society, 477, 1206

\bibitem[{{Pian} {et~al.}(2017){Pian}, {D'Avanzo}, {Benetti}, {Branchesi},
  {Brocato}, {Campana}, {Cappellaro}, {Covino}, {D'Elia}, \& {Fynbo}}]{pian}
{Pian}, E., {D'Avanzo}, P., {Benetti}, S., {et~al.} 2017, \nat, 551, 67

\bibitem[{Reifarth(2010)}]{rene}
Reifarth, R. 2010, Journal of Physics: Conference Series, 202, 012022

\bibitem[{{Roederer} \& {Lawler}(2012)}]{roeder}
{Roederer}, I.~U. \& {Lawler}, J.~E. 2012, The Astrophysical Journal, 750, 76

\bibitem[{{Siegel} {et~al.}(2019){Siegel}, {Barnes}, \& {Metzger}}]{siegel}
{Siegel}, D.~M., {Barnes}, J., \& {Metzger}, B.~D. 2019, \nat, 569, 241

\bibitem[{{Siqueira Mello} {et~al.}(2012){Siqueira Mello}, {Barbuy}, {Spite},
  \& {Spite}}]{sique}
{Siqueira Mello}, C., {Barbuy}, B., {Spite}, M., \& {Spite}, F. 2012, \aap,
  548, A42

\bibitem[{{Smartt} {et~al.}(2017){Smartt}, {Chen}, {Jerkstrand}, {Coughlin},
  {Kankare}, {Sim}, {Fraser}, {Inserra}, {Maguire}, \& {Chambers}}]{smartt}
{Smartt}, S.~J., {Chen}, T.~W., {Jerkstrand}, A., {et~al.} 2017, \nat, 551, 75

\bibitem[{{Sneden} \& {Cowan}(2003)}]{sneden03}
{Sneden}, C. \& {Cowan}, J.~J. 2003, Science, 299, 70

\bibitem[{{Sneden} {et~al.}(2008){Sneden}, {Cowan}, \& {Gallino}}]{sneden08}
{Sneden}, C., {Cowan}, J.~J., \& {Gallino}, R. 2008, \araa, 46, 241

\bibitem[{{Thielemann} {et~al.}(2011){Thielemann}, {Arcones}, {K{\"a}ppeli},
  {Liebend{\"o}rfer}, {Rauscher}, {Winteler}, {Fr{\"o}hlich}, {Dillmann},
  {Fischer}, {Martinez-Pinedo}, {Langanke}, {Farouqi}, {Kratz}, {Panov}, \&
  {Korneev}}]{thiele11}
{Thielemann}, F.~K., {Arcones}, A., {K{\"a}ppeli}, R., {et~al.} 2011, Progress
  in Particle and Nuclear Physics, 66, 346

\bibitem[{{Thielemann} {et~al.}(2018){Thielemann}, {Isern}, {Perego}, \& {von
  Ballmoos}}]{thiele18}
{Thielemann}, F.-K., {Isern}, J., {Perego}, A., \& {von Ballmoos}, P. 2018,
  \ssr, 214, 62

\bibitem[{{Woosley} {et~al.}(1994){Woosley}, {Wilson}, {Mathews}, {Hoffman}, \&
  {Meyer}}]{woosley1}
{Woosley}, S.~E., {Wilson}, J.~R., {Mathews}, G.~J., {Hoffman}, R.~D., \&
  {Meyer}, B.~S. 1994, \apj, 433, 229

\bibitem[{{Yanny} {et~al.}(2009){Yanny}, {Rockosi}, {Newberg}, {Knapp},
  {Adelman-McCarthy}, {Alcorn}, {Allam}, {Allende Prieto}, {An}, \&
  {Anderson}}]{segue}
{Yanny}, B., {Rockosi}, C., {Newberg}, H.~J., {et~al.} 2009, \aj, 137, 4377

\end{thebibliography}

\appendix 

\onecolumn
\section{SDSS/SEGUE stars potentially rich in r-process elements}
\label{App:stars}

\begin{table*}[!htb]
\small
\centering
\caption{\label{tab:stars}Spectroscopic identifier, spectral subclass and metallicity of stars with likely enhancement of neutron-rich elements that are located in SEGUE plates listed in Table~\ref{tab:results}. Secondary candidates (i.e.,\ lacking both the \ion{Eu}{I} 4594 \AA\ and the \ion{Eu}{II} 4129 \AA\ lines) are identified with an asterisk.}
\begin{tabular}{l c l c l c l}
\hline\hline
\addlinespace[0.05cm]
\textrm{Plate ID}&
\textrm{Cluster ID}&
\multicolumn{1}{c}{\textrm{{\texttt{SpecobjID}}}}&
\textrm{Subclass}&
\multicolumn{1}{c}{\textrm{[Fe/H]}}&
\textrm{$l$}&
\textrm{Spectral lines detected}
\\
\hline
\addlinespace[0.05cm]
\multirow{3}{*}{3304} & \multirow{3}{*}{--} & 3719984370237415424 & F5 & \multicolumn{1}{c}{--} & 3 & \ion{Eu}{I}$\lambda$4594; \ion{Eu}{II}$\lambda$4129,4205 \\ && 3719973924876951552 & F5 & $-1.240$ & 3 & \ion{Eu}{I}$\lambda$4594; \ion{Eu}{II}$\lambda$4129,4205\\ && 3720148197469954048 & F5 & $-1.668$ & 3 & \ion{Eu}{I}$\lambda$4594,4627,4661\\ \hline
\addlinespace[0.05cm]
\multirow{6}{*}{2667} &\multirow{6}{*}{M67} & 3002942521685941248 & F2 & $-1.519$ & 3 & \ion{Eu}{I}$\lambda$4594,4627; \ion{Eu}{II}$\lambda$4205\\ && 3002810305412701184 & F5 & $-0.841$ & 3 & \ion{Eu}{I}$\lambda$4661; \ion{Eu}{II}$\lambda$4129,4205\\&& 3002885072203389952 & F9 & $-0.449$ & 3 & \ion{Eu}{I}$\lambda$4594,4627,4661\\&& 3002824049308048384 & G2 & $-0.463$ & 3 & \ion{Eu}{I}$\lambda$4594,4661; \ion{Eu}{II}$\lambda$4205\\&& 3002777320063867904 & F9 & $-0.774$ & 4 & \ion{Eu}{I}$\lambda$4627,4661; \ion{Eu}{II}$\lambda$4129,4205\\&& 3002878750011530240 & F9 & $-0.841$ & 4 & \ion{Eu}{I}$\lambda$4594,4661; \ion{Eu}{II}$\lambda$4129,4205  \\\hline
\addlinespace[0.05cm]
\multirow{5}{*}{1961} & \multirow{5}{*}{M2} & 2208063770381622272 & F9 & $-0.519$ & 3 & \ion{Eu}{I}$\lambda$4594; \ion{Eu}{II}$\lambda$4129,4205\\&& 2208000273585118208 & F9 & $-0.650$ & 3 & \ion{Eu}{I}$\lambda$4594,4627; \ion{Eu}{II}$\lambda$4205\\&& 2207981856765352960 & F9 & $-1.721$ & 3 & \ion{Eu}{I}$\lambda$4627,4661; \ion{Eu}{II}$\lambda$4129\\&& 2207959591654890496* & G2 & $-1.790$ & 3 & \ion{Eu}{I}$\lambda$4627,4661; \ion{Eu}{II}$\lambda$4205\\&& 2208003297242094592 & G2 & $-1.534$ & 4 & \ion{Eu}{I}$\lambda$4627,4661; \ion{Eu}{II}$\lambda$4129,4205\\&& 2208044803806043136 & F5 & $-2.530$ & 5 & \ion{Eu}{I}$\lambda$4594,4627,4661; \ion{Eu}{II}$\lambda$4129,4205\\ \hline
\addlinespace[0.05cm]
\multirow{10}{*}{2476} & \multirow{10}{*}{NGC5053} & 2787852478346009600* & F2 & $-2.230$ & 3 & \ion{Eu}{I}$\lambda$4627,4661; \ion{Eu}{II}$\lambda$4205\\&& 2787767541072763904 & F5 & $-1.052$ & 3 & \ion{Eu}{I}$\lambda$4594,4627,4661\\&& 2787760669125090304 & F9 & $-1.078$ & 3 & \ion{Eu}{I}$\lambda$4594,4627,4661 \\&& 2787743351816952832 & F5 & $-1.062$ & 3 & \ion{Eu}{I}$\lambda$4594,4661; \ion{Eu}{II}$\lambda$4129\\&& 2787843132497173504 & G0 & $-1.817$ & 3 & \ion{Eu}{I}$\lambda$4661; \ion{Eu}{II}$\lambda$4129,4205\\&& 2787852203468102656 & G2 & $-2.113$ & 3 & \ion{Eu}{I}$\lambda$4627,4661; \ion{Eu}{II}$\lambda$4129\\&& 2787789256427412480 & G2 & $-1.812$ & 3 & \ion{Eu}{I}$\lambda$4627,4661; \ion{Eu}{II}$\lambda$4129\\&& 2787840108840197120 & F9 & $-0.823$ & 3 & \ion{Eu}{I}$\lambda$4594,4627; \ion{Eu}{II}$\lambda$4205\\&& 2787877492235541504 & F9 & $-0.691$ & 4 & \ion{Eu}{I}$\lambda$4594,4627,4661; \ion{Eu}{II}$\lambda$4205\\&& 2787754072055323648 & G2 & $-1.198$ & 4 & \ion{Eu}{I}$\lambda$4594,4627,4661; \ion{Eu}{II}$\lambda$4205\\ \hline 
\addlinespace[0.05cm]
\multirow{18}{*}{2475} & \multirow{18}{*}{M3} & 2786624599054457856* & F9 & $-0.588$ & 3 & \ion{Eu}{I}$\lambda$4627,4661; \ion{Eu}{II}$\lambda$4205 \\&& 2786709261449796608 & F9 & \multicolumn{1}{c}{--} & 3 & \ion{Eu}{I}$\lambda$4594,4627; \ion{Eu}{II} 4205\\&& 2786609755647482880 & F9 & \multicolumn{1}{c}{--} & 3 & \ion{Eu}{I}$\lambda$4627; \ion{Eu}{II} 4129,4205\\&& 2786707062426541056 & F9 & \multicolumn{1}{c}{--} & 3 & \ion{Eu}{I}$\lambda$4594; \ion{Eu}{II}$\lambda$4129,4205\\&& 2786708436816075776 & F9 & \multicolumn{1}{c}{--} & 3 & \ion{Eu}{I}$\lambda$4594; \ion{Eu}{II}$\lambda$4129,4205\\&& 2786700190478867456* & F5 & \multicolumn{1}{c}{--} & 3 & \ion{Eu}{I}$\lambda$4627,4661; \ion{Eu}{II}$\lambda$4205\\&& 2786633944903293952 & G2 & \multicolumn{1}{c}{--} & 3 & \ion{Eu}{I}$\lambda$4594,4661; \ion{Eu}{II}$\lambda$4129\\&& 2786630646368410624 & F9 & \multicolumn{1}{c}{--} & 3 & \ion{Eu}{I}$\lambda$4594,4661; \ion{Eu}{II}$\lambda$4129\\&& 2786611954670738432 & F9 & $-1.213$ & 3 & \ion{Eu}{I}$\lambda$4594,4661; \ion{Eu}{II}$\lambda$4129\\&& 2786730152170724352 & G0 & $-1.837$ & 3 & \ion{Eu}{I}$\lambda$4594,4661; \ion{Eu}{II}$\lambda$4205\\&& 2786726029002120192 & F9 & $-1.229$ & 3 & \ion{Eu}{I}$\lambda$4594,4627,4661\\&& 2786658409037011968 & F5 & $-1.368$ & 3 & \ion{Eu}{I}$\lambda$4594,4627,4661\\&& 2786681773659102208 & F9 & $-1.475$ & 3 & \ion{Eu}{I}$\lambda$4627,4661; \ion{Eu}{II}$\lambda$4129\\&& 2786704863403285504 & F5 & $-1.615$ & 4 & \ion{Eu}{I}$\lambda$4594,4627,4661; \ion{Eu}{II}$\lambda$4129\\&& 2786753516792814592 & F2 & $-1.698$ & 4 & \ion{Eu}{I}$\lambda$4594,4627,4661; \ion{Eu}{II}$\lambda$4205 \\&& 2786720531443981312 & F2 & $-1.621$ & 4 & \ion{Eu}{I}$\lambda$4594,4627,4661; \ion{Eu}{II}$\lambda$4129 \\&& 2786716958031191040 & F5 & $-2.056$ & 4 & \ion{Eu}{I}$\lambda$4594,4627,4661; \ion{Eu}{II}$\lambda$4129\\&& 2786636418804456448 & F9 & \multicolumn{1}{c}{--} & 5 & \ion{Eu}{I}$\lambda$4594,4627,4661; \ion{Eu}{II}$\lambda$4129,4205\\ \addlinespace[0.05cm]\hline
\end{tabular}
%\addlinespace[0.1cm]
\end{table*} 

\end{document}